\documentclass[sn-mathphys,Numbered]{sn-jnl}

\usepackage{graphicx}%
\usepackage{multirow}%
\usepackage{amsmath,amssymb,amsfonts}%
\usepackage{amsthm}%
\usepackage{mathrsfs}%
\usepackage[title]{appendix}%
\usepackage{xcolor}%
\usepackage{textcomp}%
\usepackage{manyfoot}%
\usepackage{booktabs}%
\usepackage{algorithm}%
\usepackage{algorithmicx}%
\usepackage{algpseudocode}%
\usepackage{listings}%
\usepackage[normalem]{ulem}

\usepackage{natbib}
\begin{document}

\title{Addressing the problem of the LIGO-Virgo-KAGRA visibility in the scientific literature}
\date{June 2023}

\author[1,2]{\fnm{Pablo} \sur{Barneo}}

\author[3,4]{\fnm{Giuseppe} \sur{Cabras}}

\author[5]{\fnm{Pierre-Francois} \sur{Cohadon}}

\author[6]{\fnm{Livia} \sur{Conti}}

\author[7]{\fnm{Davide} \sur{Guerra}}

\author[4,8]{\fnm{Edoardo} \sur{Milotti}}

\author[9]{\fnm{Jerome} \sur{Novak}}

\author[4,8]{\fnm{Agata} \sur{Trovato}}

\author[4,8]{\fnm{Andrea} \sur{Virtuoso}}

\affil[1]{\orgdiv{Departament de Física Quàntica i Astrofísica (FQA)}, \orgname{Universitat de Barcelona (UB)}, \orgaddress{\street{c. Martí i Franquès 1}, \city{Barcelona}, \postcode{E-08028}, \country{Spain}}}

\affil[2]{\orgdiv{Institut de Ciències del Cosmos (ICCUB)}, \orgname{Universitat de Barcelona (UB)}, \orgaddress{\street{c. Martí i Franquès 1}, \city{Barcelona}, \postcode{E-08028 }}, \country{Spain}}

\affil[3]{\orgdiv{ Dipartimento di Scienze Matematiche, Informatiche e Fisiche}, \orgname{Università di Udine}, \orgaddress{\street{Via delle Scienze 206}, \city{Udine}, \postcode{I-33100}, \country{Italy}}}

\affil[4]{\orgdiv{Sezione di Trieste}, \orgname{Istituto Nazionale di Fisica Nucleare}, \orgaddress{\street{Via Valerio, 2}, \city{Trieste}, \postcode{I-34127},  \country{Italy}}}

\affil[5]{Laboratoire Kastler Brossel, Sorbonne Université, ENS - Université PSL, CNRS, Collège de France, Campus Pierre et Marie Curie, 4 place Jussieu, F75005 Paris, France}

\affil[6]{\orgdiv{Sezione di Padova}, \orgname{Istituto Nazionale di Fisica Nucleare}, \orgaddress{\street{Via Marzolo, 8}, \city{Padova}, \postcode{I-35131},  \country{Italy}}}

\affil[7]{Departamento de Astronomía y Astrofísica, Universitat de València, Dr. Moliner 50 –  46100 – Burjassot (València) – Spain}

\affil[8]{Dipartimento di Fisica, Università di Trieste,  Via Valerio 2, I-34127 Trieste, Italy}

\affil[9]{Laboratoire Univers et Théories, Observatoire de Paris, Université PSL, CNRS, F-92190 Meudon, France}

\abstract{As members of the Virgo Collaboration – one of the large scientific collaborations that explore the universe of gravitational waves together with the LIGO Scientific Collaboration and the KAGRA Collaboration – we became aware of biased citation practices that exclude Virgo, as well as KAGRA, from achievements that collectively belong to the wider LIGO/Virgo/KAGRA Collaboration. Here, we frame these practices in the context of Merton’s ``Matthew effect'', extending the reach of this well studied cognitive bias to include large international scientific collaborations. We provide qualitative evidence of its occurrence, displaying the network of links among published papers in the scientific literature related to Gravitational Wave science. We note how the keyword ``LIGO'' is linked to a much larger number of papers and variety of subjects than the keyword ``Virgo''. We support these qualitative observations with a quantitative study based on a year-long monitoring of the relevant literature, where we scan all new preprints appearing in the arXiv electronic preprint database. Over the course of one year we identified all preprints failing to assign due credits to Virgo. As a further step, we undertook positive actions by asking the authors of problematic papers to correct them. Here, we also report on a more in--depth investigation which we performed on problematic preprints that appeared in the first three months of the period under consideration, checking how frequently their authors reacted positively to our request and corrected their papers. Finally, we measure the global impact of papers classified as problematic and observe that, thanks to the changes implemented in response to our requests, the global impact (measured as the number of citations of papers which still  contain Virgo visibility issues) was halved. We conclude the paper with general considerations for future work in a wider perspective.}

\maketitle

\section{Introduction}
Science is often imagined as purely objective and rational, but being a human endeavor, it shares the same lights and shadows of all the other human enterprises. The work of Harriet Zuckerman (\cite{Zuckerman_1965},\cite{Zuckerman_1977})  and the comprehensive analysis of Robert K. Merton (ref. \cite{Merton_1996}) clearly showed that while scientists are generally viewed as peers, some scientists are more authoritative -- and ``more equal'' -- than others. This leads to a specific cognitive bias which Merton called ``the Matthew effect'' \cite{Merton_1968}) from the passage in the Gospel according to St. Matthew\footnote{The passage reads: ``For unto everyone that hath shall be given, and he shall have abundance; but from him that hath not shall be taken away even that which he hath'' (Matthew 25:29).}. Roughly speaking, this means that in the case of collaborative work or multiple simultaneous discoveries, the already famous scientists get all or most of the credit.

While the Matthew effect applied to individual scientists is well-known and has been studied for more than 50 years, only recently sociologists of science have started investigating the impact of the Matthew effect on research funding \cite{Liao_2021} and the consequent boost on productivity \cite{Zhang_2022}. In this paper we consider the specific case of LIGO, Virgo and KAGRA, where we detect a similar pattern of imbalance in the assignment of due credit by many members of the wider scientific community.

We wish to stress from the outset that this study is done in the spirit of the early work of Harriet Zuckerman, it tries to recognize the phenomenon itself -- in this case a social rather than an individual effect -- which affects large scientific collaborations, and which is different from the recently observed bias in research funding of academic organizations \cite{Liao_2021}. The study was stimulated by qualitative observations that we describe below, and which were confirmed by the citation bias that we observed in the preprint database. Although we make quantitative estimates, we do not try to model the effect as it is based on observations in just one, very specific, field of science (for modern attempts to model the Matthew effect at the single scientist level, see \cite{perc2014matthew,katchanov2023empirical}). 

In Section \ref{sec:context} we describe the scientific context in which we work and the problems associated with the observed bias. In Section \ref{sec:examples} we give a few specific examples. Next, in Section \ref{sec:size} we report about our work to detect and single out the most problematic papers. In Section \ref{sec:email} we report the actions we took to solve the issue. In Section \ref{sec:followup} we report a follow-up analysis performed on a subset of  problematic papers, 
based on the fraction of authors who corrected the  problematic text lacking proper credits to Virgo.
Finally, we frame
this biased narrative of the scientific developments in the context of the generalized ``Matthew effect'' applied to large research organizations rather than to individual scientists, and explain its appearance in the history of the LIGO, Virgo and KAGRA Collaborations. We conclude Section \ref{sec:discussion} with general remarks of interest to the wider scientific community.

\section{The scientific context}
\label{sec:context}

The first discovery of Gravitational Waves (GW) by the LIGO Scientific Collaboration and the Virgo Collaboration is a milestone of modern science: it occurred in 2015 and the result was made public in early 2016 \cite{Abbott_2016}, a century after the first prediction by Einstein \cite{einstein1916naherungsweise,rothman2018secret} and after decades of worldwide experimental effort \cite{cervantes2016brief}. After that, almost a hundred signals have been found by the LIGO Scientific, Virgo and KAGRA (LVK) Collaboration in the three observing runs of the network of the two Advanced LIGO detectors in the US and the Advanced Virgo detector in Italy \cite{GWTC-3}. Such discoveries have attracted a lot of attention and GW-based astrophysics and cosmology have become a fertile field of science.

The LIGO Scientific Collaboration consists of more than 1400 members, in representation of almost 130 institutions distributed in 19 countries worldwide (as of May 2021 \cite{LCSweb}). Presently the Virgo Collaboration consists of $\sim 800$  members, in representation of almost 150 institutions distributed in more than 15, mostly European countries: the full list of institutions to which members of the Virgo Collaboration are affiliated is available at the link in ref.\cite{Virgoweb}. The KAGRA collaboration is composed of more than 400 individuals from more than 110 institutions in 15 countries and regions around the world (as of August 2020 \cite{KAGRAweb}).

\begin{figure}[htbp!]
\centering
\includegraphics[width=1\linewidth]{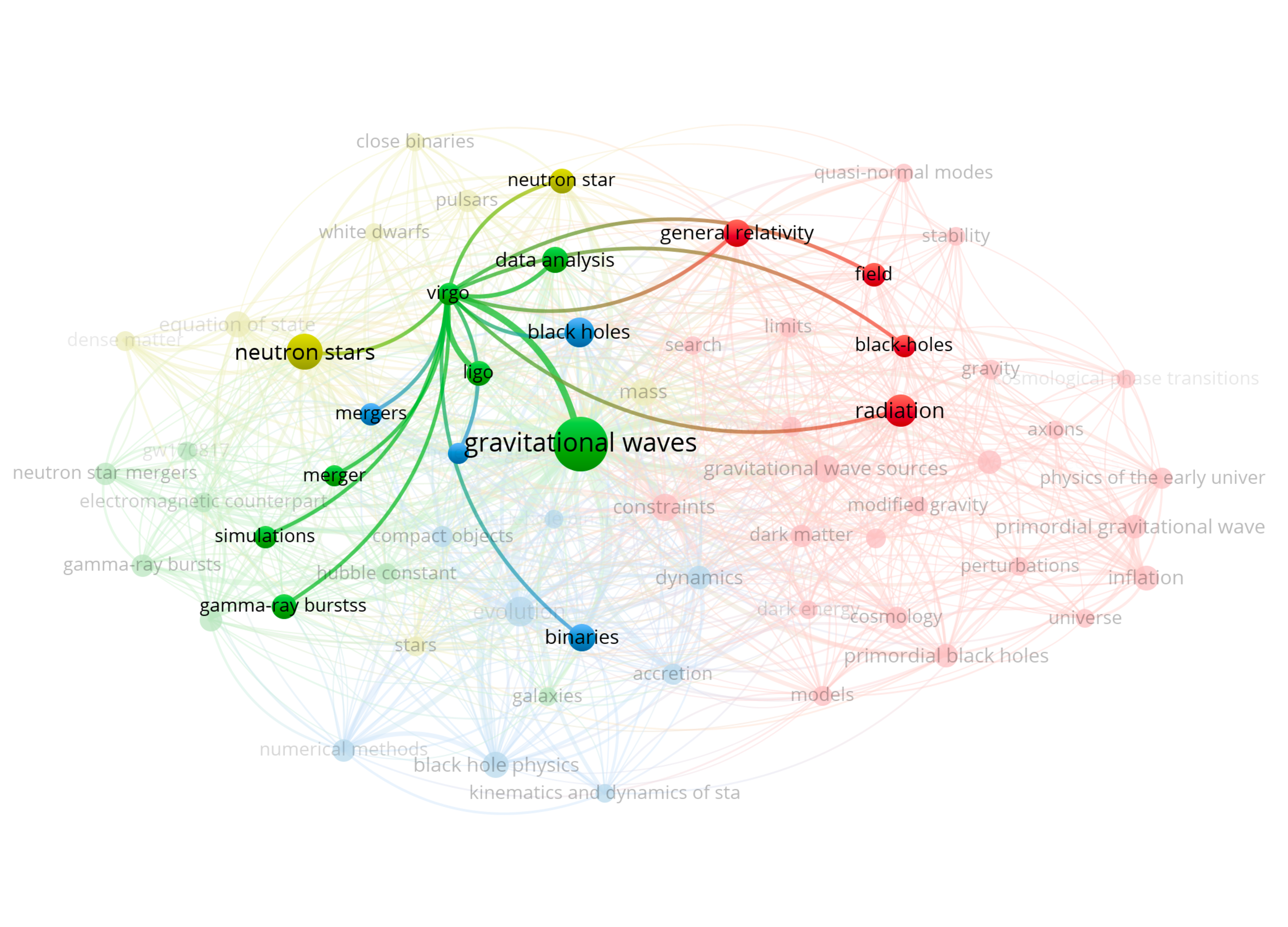}
\includegraphics[width=1\linewidth]{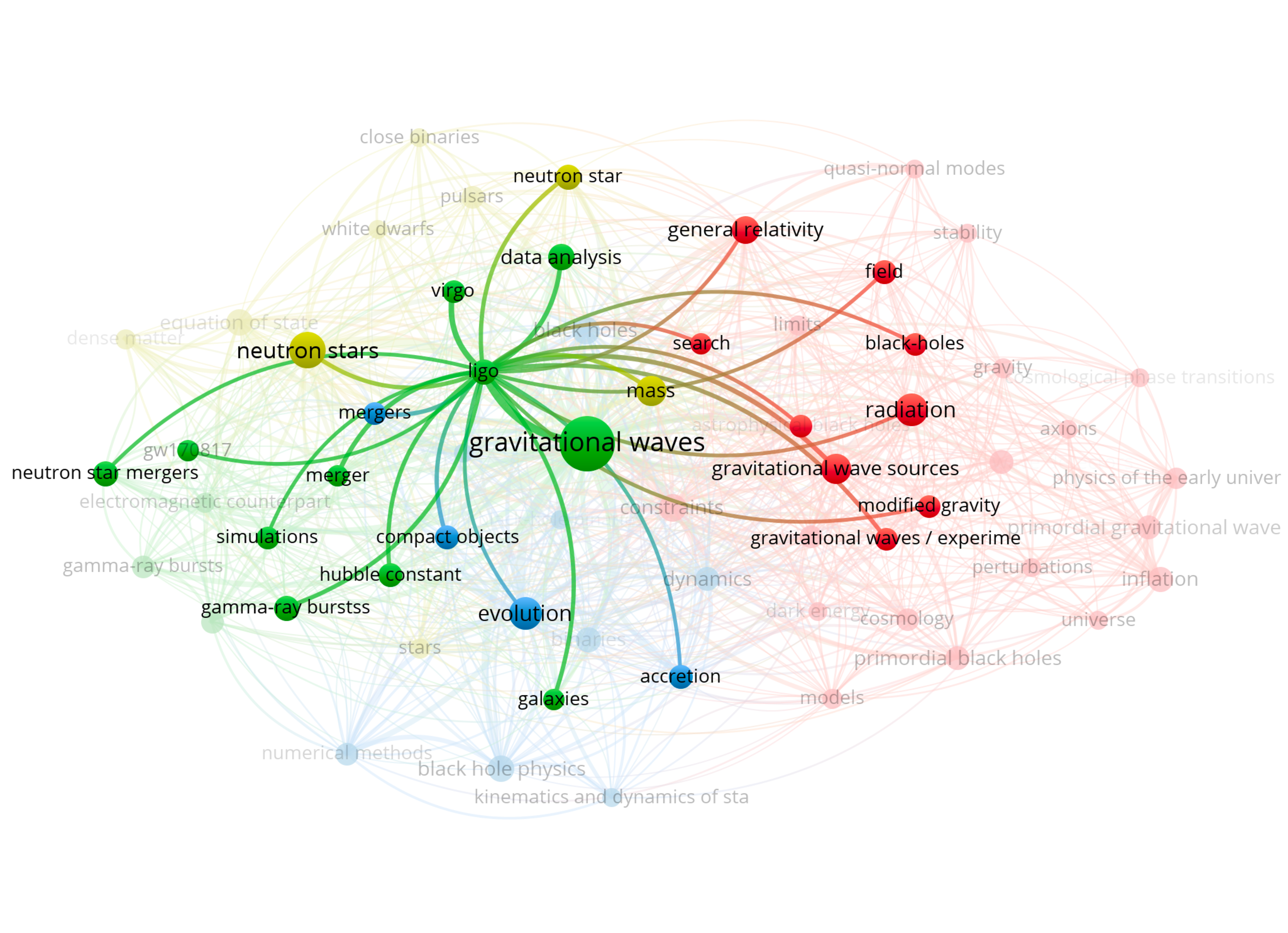}
\caption{Visualization of the network of links among published papers at the end of 2021. In both panels, the underlying network shows all the links among keywords present in the papers, obtained with VOSviewer \cite{VanEck_2022} and bibliographic data from the Web Of Science database \cite{Birkle_2020}  for papers published in the second half of 2021. The search provides a qualitative overview of the unbalance in the number of citations to LIGO and Virgo. It was carried out selecting the papers with the search string \texttt{gravitational wave OR gravitational-wave OR gravitational waves} in All Fields and keeping the 500 most recent records, which corresponds to a depth of about 6 months back in time from January 2022. The upper panel highlights the 16 direct links joining the “Virgo” keyword to other keywords, while the lower panel highlights the 26 direct links joining the ``LIGO'' keyword to other keywords. We see that ``LIGO'' is directly linked to a larger number of keywords and therefore of papers. The colors in this figure and in Fig.s \ref{fig:gw150914_links} and \ref{fig:gw170817_links} have no specific meaning, they are only used to guide the eye.}\label{fig:VirgoLigo_links}
\end{figure}

Although the LIGO and Virgo Collaborations have been acting as a single collaboration since 2007, co-authoring all of their gravitational-wave observational result papers  since 2010, we note that this is not always recognized in the wider scientific community where it is frequent to find that shared results are attributed to LIGO only (see Fig. \ref{fig:VirgoLigo_links}). This same pattern is repeating with the KAGRA Collaboration, which joined the LIGO and Virgo collaborations in 2020. 

We wish to stress that we are not highlighting an unbalanced number of citations: because of the joint authorship this cannot be the case. Rather, it is the very text of some scientific papers that does not correctly acknowledge the results of the collaborative work. In other words it is the narrative of the scientific developments that is being biased, with a potential impact on the history of science itself. So the concern here is not whether Virgo papers are cited or not, but rather if Virgo is correctly placed and associated with its science in the article’s text.

The relevance of the problem is given by the detrimental consequences it may have, both in the short and in the long term. A fair association of Virgo and now KAGRA with GW discoveries and science, along with LIGO, can affect  the work of hundreds of people working within the Virgo and KAGRA collaborations with special impact on the careers of young scientists associated with Virgo or KAGRA. This problem propagates to funding agencies which are influenced by the media mirroring the incomplete credits given by some of the problematic papers.

\medskip

\section{Specific examples}\label{sec:examples}
\begin{figure}[htbp!]
\centering
\includegraphics[width=.5\linewidth, angle=-90]{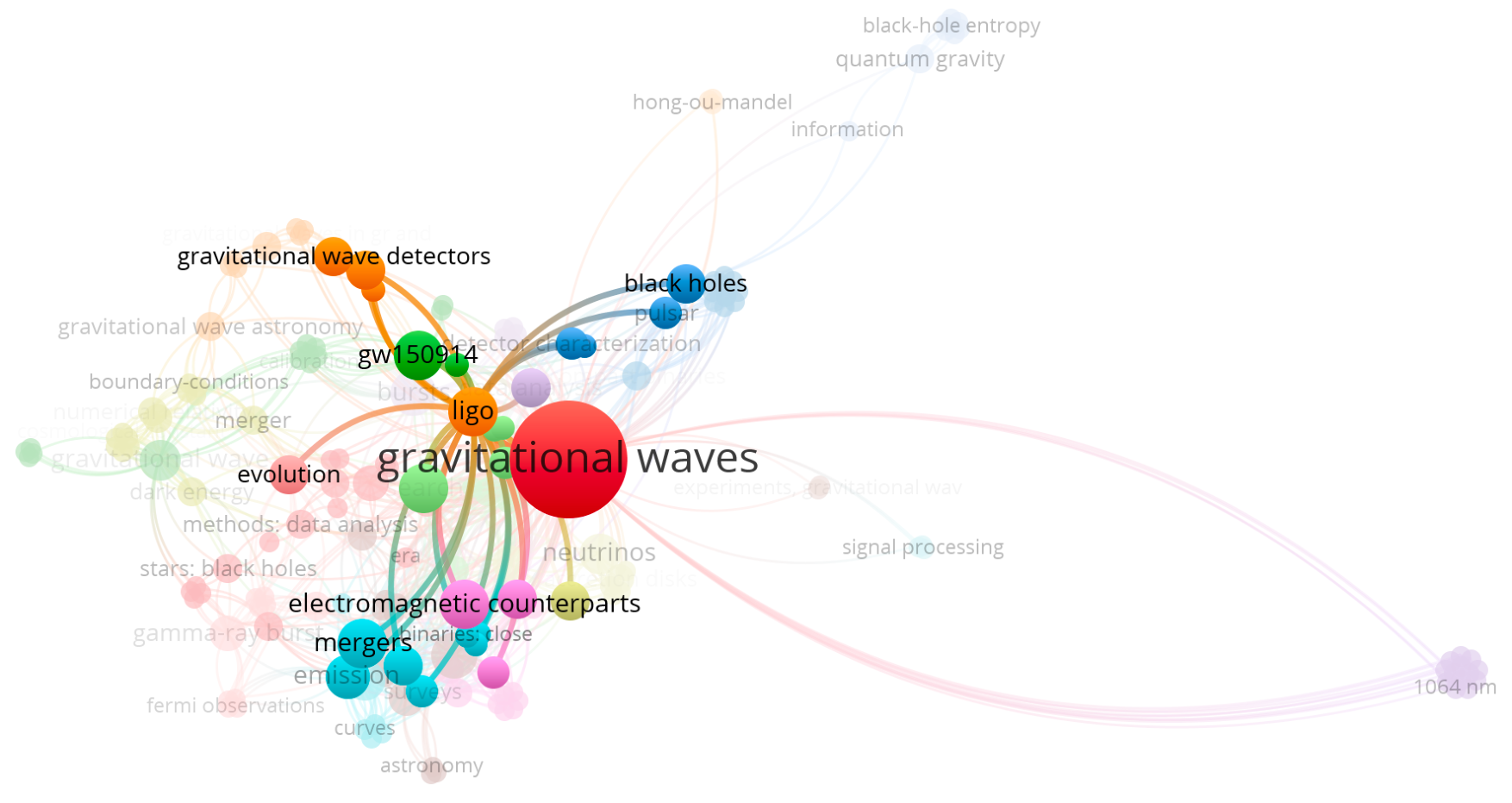}
    \caption{Visualization of the network of links among published papers dealing with the first gravitational-wave event, GW150914. The underlying network shows all the links among keywords present in the papers, obtained with VOSviewer \cite{VanEck_2022} and bibliographic data from the Web Of Science database \cite{Birkle_2020} for papers published until May 2023. The search was carried out at the end of May 2023, selecting the papers with the search string \texttt{GW150914 AND (gravitational wave OR gravitational-wave OR gravitational waves)} in All Fields, and keeping all the retrieved records.  The links to/from the ``LIGO'' keyword are highlighted. There are no links to the ``Virgo'' keyword. 
}\label{fig:gw150914_links}
\end{figure}

\begin{figure}[htbp!]
\centering
\includegraphics[width=.7\linewidth, angle=-90]{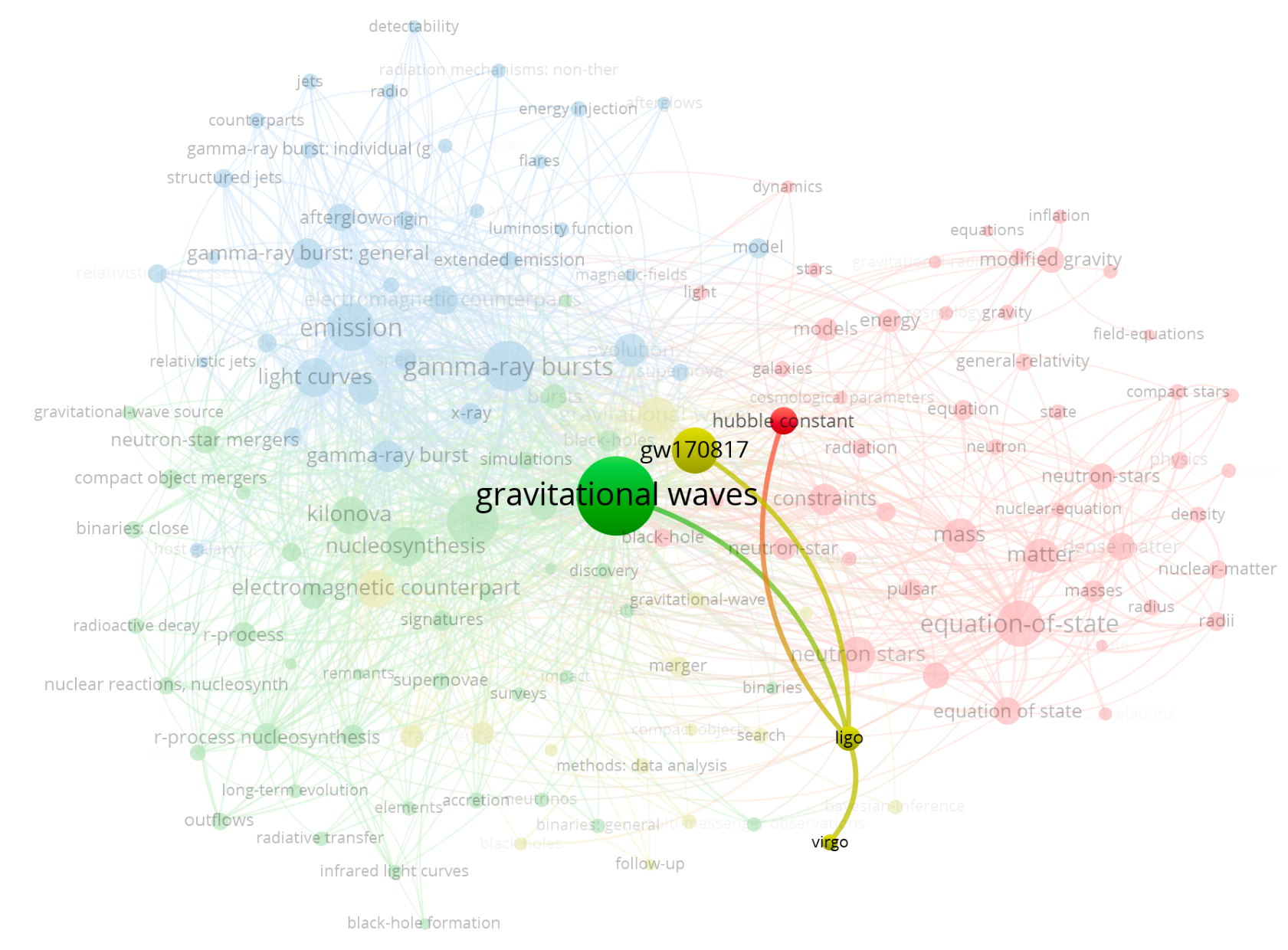}
    \caption{ Visualization of the network of links among published papers dealing with the first binary neutron star coalescence, GW170817. The underlying network shows all the links among keywords present in the papers, obtained with VOSviewer \cite{VanEck_2022} and bibliographic data from the Web Of Science database \cite{Birkle_2020} for papers published until May 2023. The search was carried out at the end of May 2023, selecting the papers with the search string \texttt{GW170817 AND (gravitational wave OR gravitational-wave OR gravitational waves)} in All Fields, and keeping all the retrieved records. The links to/from the ``LIGO'' keyword are highlighted. Note that the ``Virgo'' keyword only links to the ``LIGO'' keyword. GW170817 was the first gravitational-wave event with an electromagnetic counterpart, and was observed by more than 60 observatories, with observations spanning the whole spectrum, from radio waves to gamma rays \cite{Abbott_2017b}: this explains the large number of keywords that are not directly linked to the gravitational-wave observatories.}\label{fig:gw170817_links}
\end{figure}

In this section we consider in detail specific examples of the issues we are discussing. In particular, we refer to a few papers that were published before we started our systematic work, detailed in the next section. We want to stress that we are not pointing at the authors of these specific papers, but just using their texts to extract some examples that we consider typical of a more general habit. We find that the issues can be broadly divided into three categories: 
\begin{enumerate}
    \item A very frequent issue is related to the first event GW150914, which may be due to the fact that only the Advanced LIGO detectors were online at the time of the first detection. Indeed, one should distinguish between the data containing the signal that on this occasion were acquired by the two Advanced LIGO detectors only and the discovery of that signal, which was the result of the collaborative effort of the scientists from both LIGO and Virgo, who eventually co-authored the ground-breaking GW150914 discovery paper that is cited by all authors who fail to mention Virgo \cite{Abbott_2016}. An instance of LIGO-only citation  is offered by the very first sentence of a preprint which appeared in 2020, and which has already been cited more than 80 times (as of June 2023): ``The first detection of gravitational waves from a binary black hole merger by LIGO opened a new door to explore cosmology.'' Another typical example comes from a paper published in Physical Review Letters in 2016, a seminal paper with a citation count greater than 700 (as of June 2023). Here, the bias we are discussing is evident from the very title of the paper where only LIGO is mentioned, and also in the paper body which mentions ``... the two $\sim$30M$_\odot$ black holes detected by LIGO ...'' A third example paper was published in 2020 in the journal Classical and Quantum Gravity and has to date more than 80 citations (as of June 2023): in this paper the authors state that ``... On 14 September 2015 the twin LIGO detectors made the first observation of a GW signal ...''.  Figure \ref{fig:gw150914_links} displays the widespread extent of this problem: among papers containing the keywords ``gravitational waves'' and ``GW150914'' we find no direct link to the ``Virgo'' keyword while the ``LIGO'' keyword is well represented. 
    
    \item Another frequent case is related to how the discovery of GW170817 \cite{Abbott_2017} is reported. In a paper published in 2021 in Physical Review D, which has already collected 19 citations (as of June 2023), the authors write: ``The GW170817 event detected by LIGO is the first confirmed merger event of compact stars.'' Again, in a paper published in 2022 in the Astrophysical Journal we read `` ..the kilonova associated with the gravitational wave signal from an NSM detected by LIGO, GW170817.'' Unlike the previous case, at the time of GW170817 all three detectors in the network were operating, i.e. the two Advanced LIGO and the Advanced Virgo detector, and the combination of the three was especially important because it provided a precise sky localization that allowed multiple multimessenger observations (for an introduction to the new, exciting field of multimessenger astronomy with gravitational waves, see, e.g., \cite{branchesi2016multi}). While the signal was actually observed only in the data streams of the two LIGOs, because of the combined effect of antenna pattern and sensitivity, the sky localization was made possible by the use of data from all three detectors. This was the key ingredient for the success of the electromagnetic followup campaign, which made possible the kilonova discovery referred to in the paper.  Figure \ref{fig:gw170817_links} is a graphical demonstration of this issue: among the papers containing the keywords ``gravitational waves'' and ``GW170817''  we find 33 links to ``LIGO'' with respect to just 11 to ``Virgo''. 
    
    \item Finally, we find the extreme case where the overall science results are attributed to LIGO only. An example of this comes from a recent paper, published in 2022 in the Astrophysical Journal, where they reference ``the current LIGO rates'' and ``the LIGO detections''. 
\end{enumerate}
Finally, we note that projections of future developments/achievements in this field are generally referred to or labeled only as ``LIGO''.

\section{Statistical studies}\label{sec:size}
The observation of this biased narrative in a steadily growing number of papers,  prompted us to implement an approach to systematically detect it, and if possible correct it. We decided to focus our attention on the preprints announced every day in the arXiv repository \cite{arxiv}: in particular we consider preprints posted in the subjects General Relativity and Quantum Cosmology (labelled as 'gr-qc') and/or Astrophysics (labelled as 'astro-ph'). This choice covers the broader scientific community that is potentially interested in the work of the LVK collaboration. By targeting preprints rather than published papers we have the opportunity to contact the authors and fix the issues before a paper is accepted for publication.

In detail, we run an automated script that processes all preprints appearing daily in the repository and loosely selects those which are likely to contain issues. The script downloads and scans the pdf version of each paper listed in the daily digest, counting the number of times the strings ``LIGO'' and ``Virgo'' appear in the text (a case insensitive search) and finding the lines where the two strings are unmatched. Whenever this occurs, the script issues an alert, and all the alerts issued on a given date are collectively sent via email to a group of dedicated Virgo scientists. A few exceptions are granted in the script to avoid issuing unnecessary alerts, such as in the cases of internet links, affiliations, bibliography.

On a weekly basis, papers causing alerts are checked manually and those found to contain major issues are marked as problematic, according to shared general guidelines – although individual perceptions may still play a role. We carried out these checks systematically from 17 January 2022 until 27 January 2023, to sample a sizable fraction of all scientists active in our expanded community, i.e., scientists working in the fields covered by the above mentioned categories of the preprint repository.

\begin{figure}[!htt]
    \centering
    \includegraphics[width=1.\linewidth]{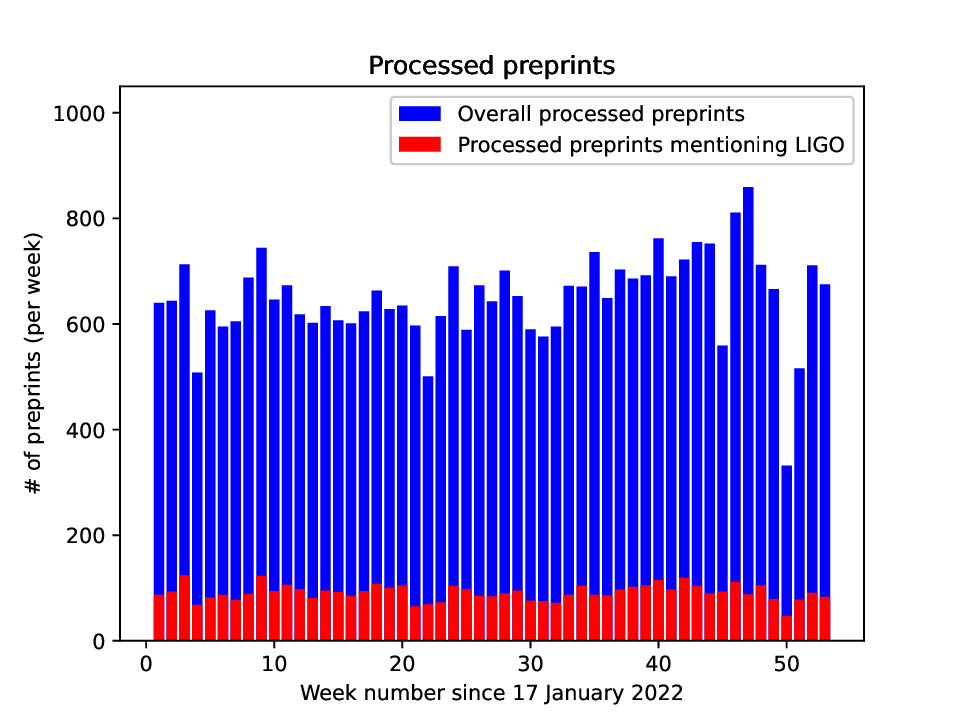}
    \caption{Total number of papers in the arXiv preprint database \cite{arxiv} in the categories 'gr-qc' and 'astro-ph'(blue) and the subset of those papers mentioning LIGO (about 14$\%$, red), on a weekly basis from 17 January 2022. Both distributions over time are roughly constant. }\label{fig:allvstime}
\end{figure}

In this period the script processed a total of 34467 papers: 4842 of them (i.e. 14$\%$) are found to contain the word ``LIGO'' (case insensitive). Fig. \ref{fig:allvstime}  shows the distribution over time of the number of processed preprints and of those containing the word ``LIGO''. The script raised an alert in 1448 cases. These are the papers that are manually inspected  on a weekly basis: those that are found to contain true major issues, similar to those in sec. \ref{sec:examples}, are marked as problematic. On average  we have manually inspected 27 papers per week and found 8 problematic papers per week. So the papers showing issues correspond to about 9$\%$ of all papers containing the word ``LIGO''.

\begin{figure}[!htbp]
    \centering
     \includegraphics[width=1.\linewidth]{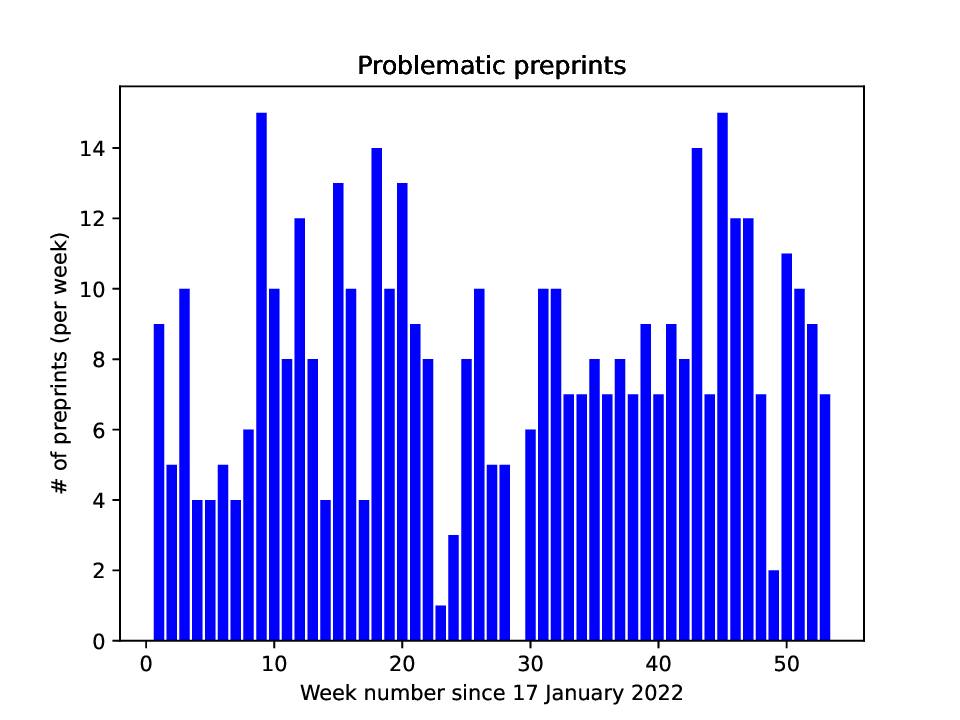}
    \caption{Number of problematic preprints over time, on a weekly basis from 17 January 2022. On average about 8 papers per week are found to contain major issues.}\label{fig:critical_abs}
\end{figure}

\begin{figure}[!htbp]
    \centering
    \includegraphics[width=1.\linewidth]{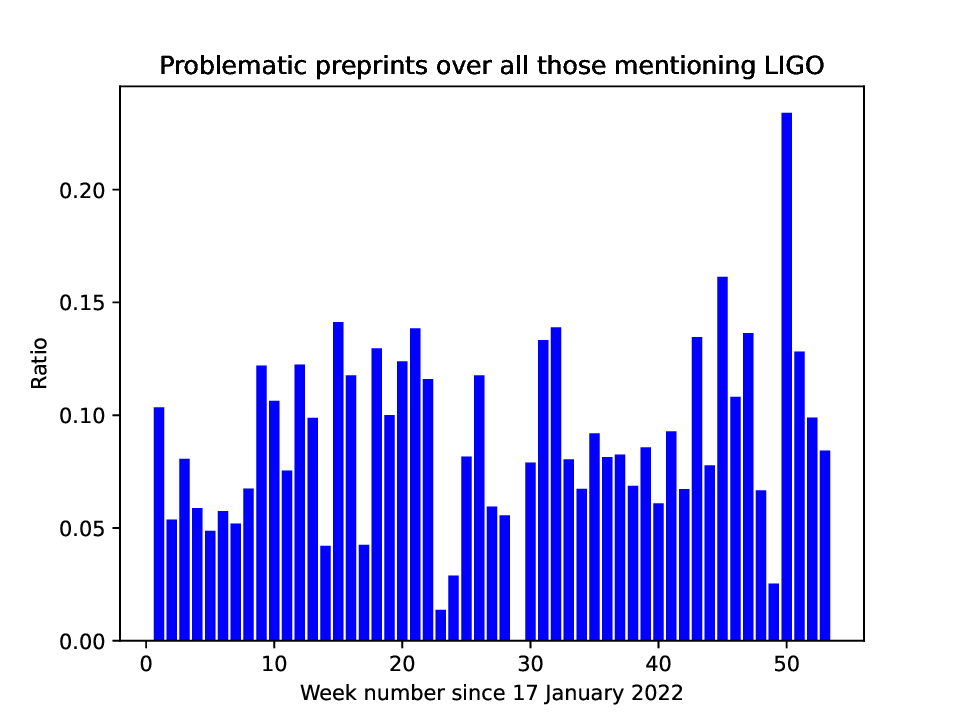}
    \caption{Fraction of problematic preprints over papers containing the word ``LIGO'' (case insensitive), on a weekly basis from 17 January 2022.  On average 9$\%$ of all papers containing the word ``LIGO'' are found to contain major issues. }\label{fig:crifical_fraction}
\end{figure}

Figure \ref{fig:critical_abs}  shows the distribution over time of the problematic papers while Figure \ref{fig:crifical_fraction}  shows the relative number of  problematic papers with respect to all papers containing the word ``LIGO'', shown on a weekly basis starting from 17 January 2022.

\section{Email exchanges}\label{sec:email}
Soon after starting the monitoring process, we actively tackled the issue. We started  sending e-mail messages to the (first) corresponding author (or to the preprint submitter when this is not available) of each preprint deemed as  problematic, as explained in sec. \ref{sec:size}. The message text is standard except for the quotes from the preprint displaying the issues we identified. The message is sent by two members of the group who also act as Virgo Outreach coordinator and co-coordinator, and sign the message specifying their role in the Collaboration. We sent messages to the authors of all problematic papers corresponding to a given week within a few hours’ interval. The delay between the appearance of the preprint in the repository and the sending of the emails ranged from a few days up to about 1 month.

We carried out this procedure for about one year: from 17 January 2022 to 6 January 2023. The choice of this period of time was dictated by the need to consider a time span long enough to sample a sizable fraction of the scientists active in our enlarged community and yet still manageable by our workforce. On one hand we aimed at fixing the problematic preprints and on the other hand to gently convince both the authors of the problematic papers and their readers, i.e. the scientific community at large, of the necessity of a correct scientific narrative.

Over the entire period under analysis, we received replies from 48$\%$ of the recipients of our e-mail messages. The vast majority expressed understanding and their willingness to edit the preprints at the next occasion. We have not checked systematically if the requested (and often promised) changes were actually implemented in the final version of the papers. However, as detailed in the following section, for the papers of the initial period we did perform a more in-depth analysis.

\section{Follow up}\label{sec:followup}

To assess the effectiveness of our actions we checked the scientific impact of problematic papers. To this end, we carried out a follow-up analysis focused on the first three months of monitoring, i.e. from 17 January 2022 until 22 April 2022. For each problematic preprint of this period we recorded the number of citations one year after their appearance on arXiv on both Google Scholar and Web of Science: while Google Scholar indexes also  non-journal sources, including theses, books, conference papers, and unpublished materials, as well as non-scholarly sources such as course readings lists and promotional pages, Web of Science considers published, peer-reviewed papers only.
This reflects in the fact that Google Scholar finds significantly more citations than Web of Science \cite{MARTINMARTIN}.

We analysed a total number of 101 problematic preprints, checking the latest available version of the preprint or the paper, if published in the meanwhile. Of 101 publications, 51 had no citations on Web of Science, meaning that the corresponding papers were not yet published, or eventually that the corresponding papers had no citations yet from published papers. For the sake of simplicity, from now on we'll refer to the publications considered one year after the appearance on arXiv of the corresponding preprints as ``papers'', without distinguish between published and unpublished (at the time of our analysis) preprints.

\begin{figure}[!htbp]
    \centering
    \includegraphics[width=1.\linewidth]{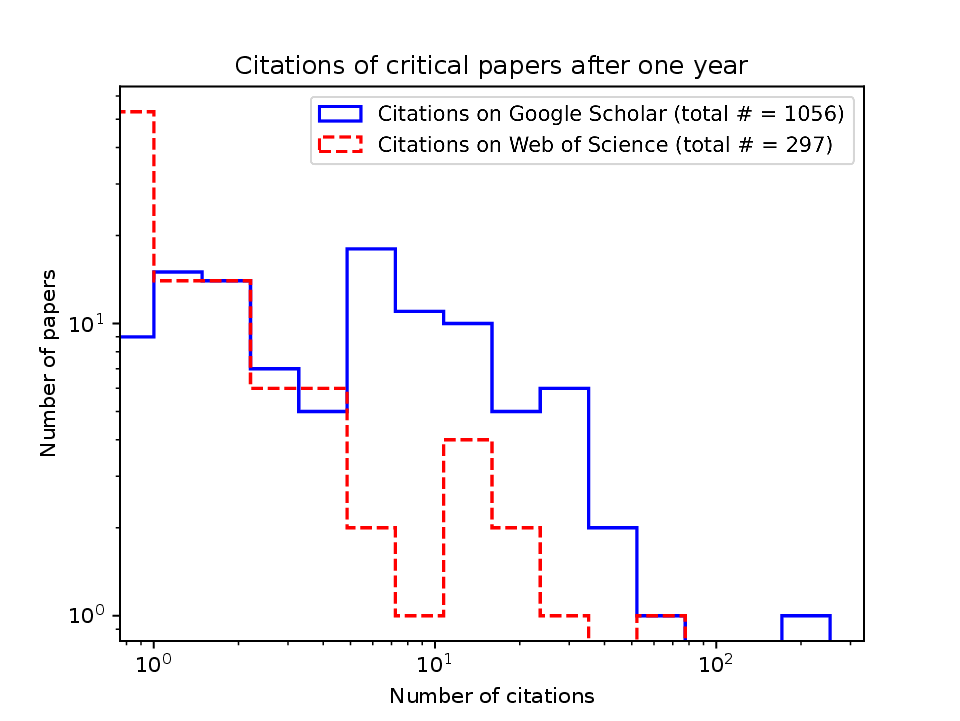}
    \caption{Citations of problematic papers in Google Scholar (blue) and in Web of Science (red) one year after their appearance on the arXiv preprint database \cite{arxiv}. In addition to the larger citation counts found by Google Scholar, which includes preprints as well as published papers, we observe that a few papers have already attracted tens of citations and may have a sizable impact on the scientific community. }\label{fig:citation_crit_google}
\end{figure}

\begin{figure}[!htbp]
    \centering
    \includegraphics[width=1.\linewidth]{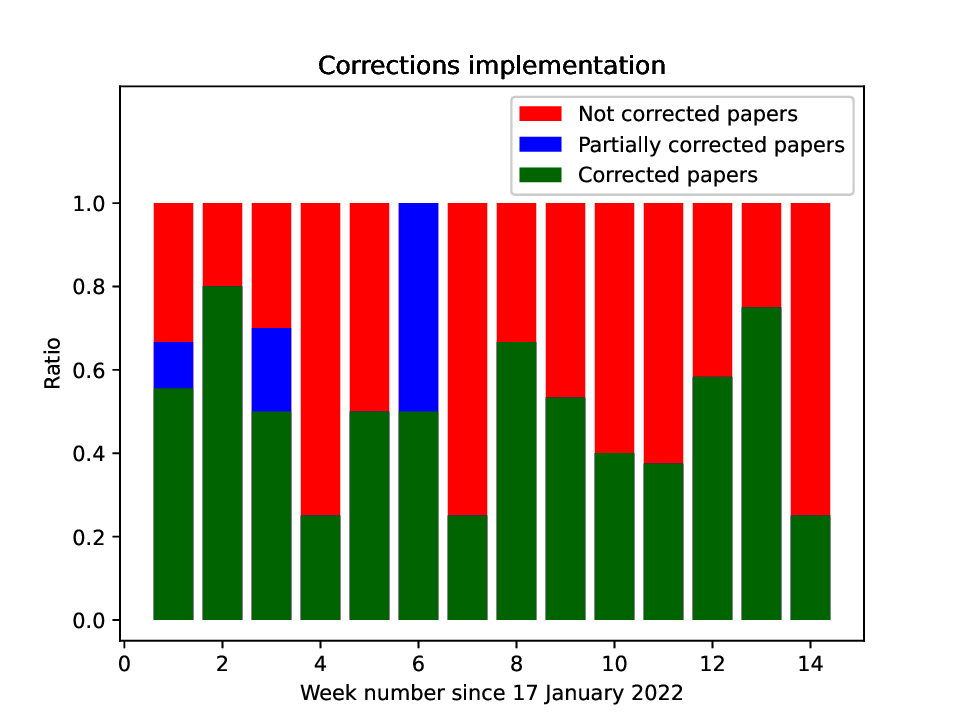}
    \caption{Fraction of papers that appeared in the arXiv preprint database \cite{arxiv} in the first three months of 2022 and which implemented the requested corrections: out of 101 papers, 52 of them ($\sim 50\%$, shown in green) have been fully corrected  following our suggestions (shown in green), in 45 cases ($\sim 45\%$, shown in red) no corrections have been implemented and in 4 cases ($\sim 5\%$, shown in blue) only some of the corrections have been are implemented. There is no clear trend which may indicate that the community is changing habits.
    }\label{fig:corr}
\end{figure}

The results are shown in Figures \ref{fig:citation_crit_google}  and \ref{fig:corr} . Figure \ref{fig:citation_crit_google}  shows histograms displaying the distribution of citations of  problematic papers with data from Google Scholar and Web of Science. If these papers had been written by a single author they would have brought his/her H-index from 0 to 16 after one year, considering the Google Scholar citations. However, it is also worth noticing that there are a few papers with more than 50 citations, and so with a considerable impact on the scientific community.

Figure \ref{fig:corr} , shows our weekly rate of success in correcting the wrong narrative. Among the overall 101 problematic papers we followed-up, we found that 52 implemented our corrections, 45 that did not, and interestingly in some cases the authors corrected only a subset of the problematic lines, leaving others uncorrected: we’ll refer to these papers (4 in all) as “partially corrected”.

\begin{figure}[!htbp]
    \centering
\includegraphics[width=.8\linewidth]{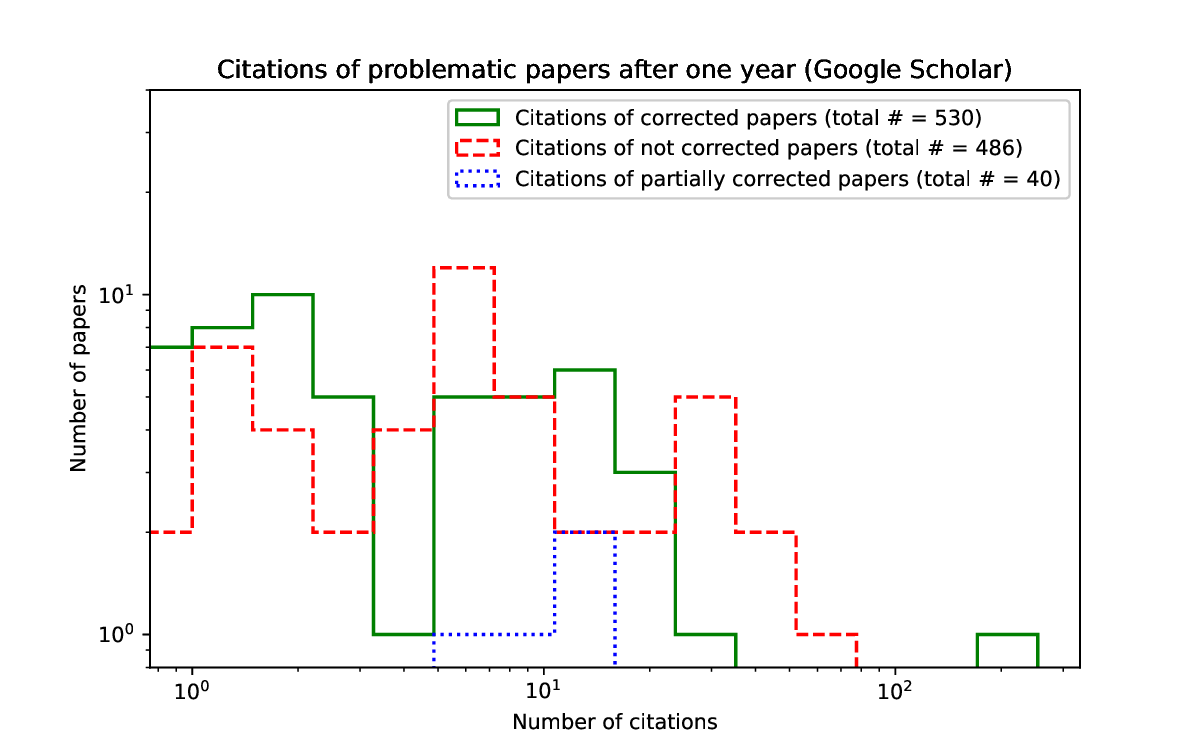}
\includegraphics[width=.8\linewidth]{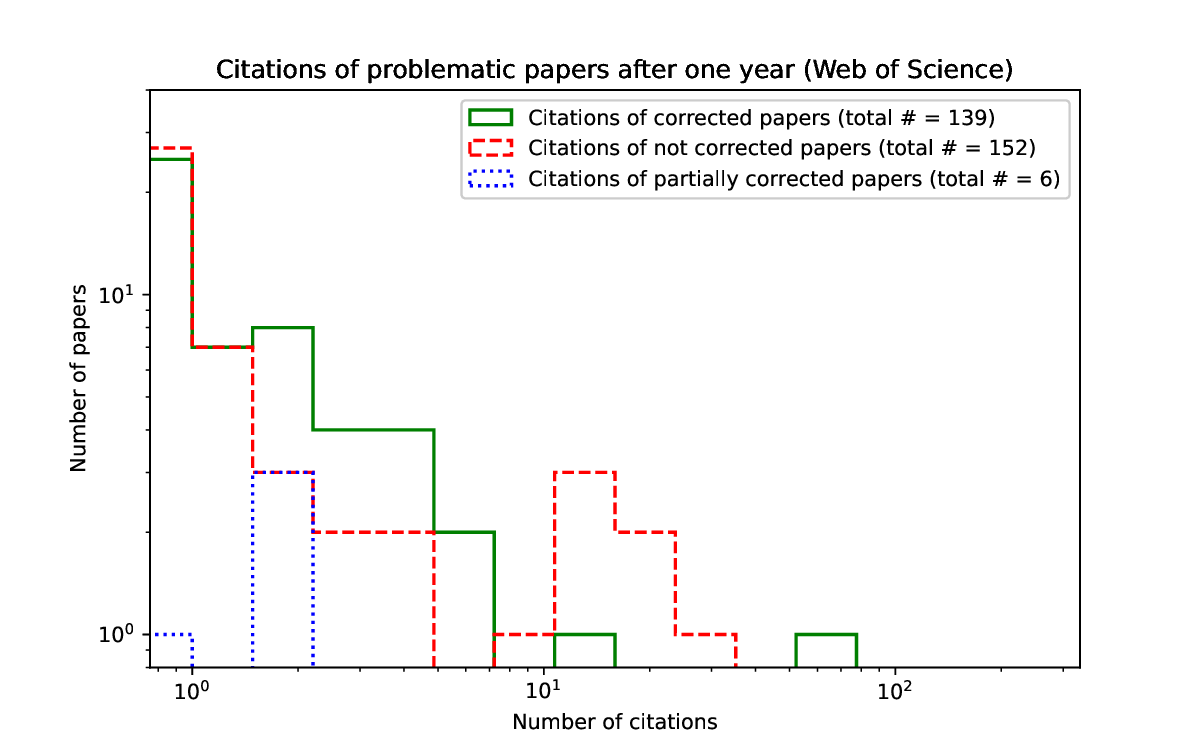}
\caption{ Citations of problematic papers one year after their appearance in the arXiv preprint database \cite{arxiv}, from Google Scholar (upper panel) and Web of Science (lower panel). Here, we aggregate the papers in three categories: fully corrected papers (shown in green), uncorrected papers (shown in red) and partially corrected papers (shown in blue). In both cases we do not observe any significant difference about the number of citations of corrected vs not corrected papers. Hence our work was effective in halving the impact of papers whose preprints were found to contain major issues. As already noticed, the citations counts provided by 
Google Scholar are generally larger with respect to Web of Science.}\label{fig:citation_histo}
\end{figure}

Next, we tried to evaluate the impact of papers based on their correction status (see Fig. \ref{fig:citation_histo} ). We found that the total number of citations for corrected papers is 530 on Google Scholar and 139 on Web of Science, for partially corrected papers 40 on Google Scholar and 6 on Web of Science, for uncorrected papers 486 on Google Scholar and 152 on Web of Science. Assuming that the number of citations in time follows a roughly Poisson distribution, this means that the difference between corrected and uncorrected papers is about 1.4 standard deviations on Google Scholar and about 1.3 on Web of Science: we conclude that both differences are not significant and that our action split the impact roughly into two equal parts, i.e., we were able to halve the impact of originally problematic papers.

\section{Discussion}\label{sec:discussion}
We find that our actions did partially correct the situation that we lamented in the introduction, as some of the papers are indeed fixed. However, our knowledge of the social dynamics of the particular cognitive bias that we try to correct is still quite poor. In particular, we find no actual difference between the impact of authors who gave us a positive response and of the authors who declined to respond and act on their papers. We also note that here we have focused on issues related to the visibility of the Virgo Collaboration. However the LIGO Scientific Collaboration and Virgo Collaboration have been joined in the beginning of 2021 by the KAGRA Collaboration in co-authoring observational results from the full third observational period O3. Hence it would be interesting to extend the study of cognitive bias by considering KAGRA as well.

There still remain unanswered questions that we briefly mention here.
We could not detect any positive evolution in the collective response of the authors of problematic papers, which is likely to have a very long time constant. To evaluate a possible long-term impact of this work we should repeat it, covering a different period of time. On the other hand we anticipate that it would be very difficult to disentangle any impact of this work from other influences that may affect this field. Still, the detection and measurement of the value of the time constant would be an important result that would shed light on socially accepted habits in this  field of science. Similar studies in other fields of knowledge could yield different values, that could help characterize the different scientific communities. 
It would also be interesting to analyze the affiliations of authors of problematic papers and check whether they match the worldwide production of Physics papers: we decided not to proceed in this direction as it would require an arbitrary weighting in the common case of multiple affiliations or of authors affiliated to institutions of different countries. However, this association could be revived -- and not just in Physics -- if sociologists of Science could produce reasonable criteria to correlate citation habits with the geographic distribution of scientists.
As a final perspective which goes much beyond the limited scope of our study, we note that our work is only partially based on citation counts, and that our results required a partially automated check of the very text of many individual papers. This was made possible by the ability to scan large amounts of text in a very short time. Further developments would certainly require a deeper examination of each text, but in this age of Artificial Intelligence, we can look forward to developments that can potentially go much further than the simple citation counts to evaluate the specific features of a scientific work. 

\section{Supplementary material}\label{sec:supmat}
We make available at the public repository in ref.\cite{supmat} all data and software used for the work reported in this paper. In particular we make available:
\begin{itemize}
    \item the result of searches in Web Of Science, used to produce figures \ref{fig:VirgoLigo_links}, \ref{fig:gw150914_links}, and \ref{fig:gw170817_links} using VOSviewer
\item the script used to scrutinize the papers appearing dayly on arXiv in the subjects General Relativity and Quantum Cosmology (labelled as gr-qc) and/or Astrophysics (labelled as astro-ph)
\item the full collection of automated e-mails sent daily to the authors of this paper and which contain the collectively alerts of the day and some statistics
\item the spreadsheet (in both Excel and LibreOffice formats) containing the results of the preprint scrutinity, grouped in weekly bunches (see sec. \ref{sec:size}) as well the data related to the follow-up analysis described in sec. \ref{sec:followup}. 
\end{itemize}

\bibliography{biblio}

\end{document}